\documentclass[sigconf, 10pt,nonacm]{acmart}

\usepackage{booktabs} 
\usepackage{subcaption}
\usepackage{balance}

 \newcommand{\edit}[1]{{\leavevmode\color{black}#1}}

\begin{document}
\title{DiOS - An Extended Reality Operating System for the Metaverse}

\author{Tristan Braud}
\affiliation{%
  \institution{Hong Kong University of Science and Technology}
    \country{Hong Kong}
}
\email{braudt@ust.hk}

\author{Lik-Hang Lee}
\affiliation{%
  \institution{
Korea Advanced Institute of Science and Technology}
    \country{South Korea}
}
\email{likhang.lee@kaist.ac.kr}

\author{Ahmad Alhilal}
\affiliation{%
  \institution{Hong Kong University of Science and Technology}
    \country{Hong Kong}
}
\email{aalhilal@ust.hk}

\author{Carlos Bermejo Fernández}
\affiliation{%
  \institution{Hong Kong University of Science and Technology}
    \country{Hong Kong}
}
\email{cbf@ust.hk}

\author{Pan Hui}
\affiliation{%
  \institution{Hong Kong University of Science and Technology, Hong Kong}
    \country{}
}
\affiliation{%
  \institution{University of Helsinki, Finland}
    \country{}
}
\email{panhui@ust.hk}

\renewcommand{\shortauthors}{T. Braud et al.}

\begin{abstract}
Driven by the recent improvements in device and networks capabilities, Extended Reality (XR) is becoming more pervasive; industry and academia alike envision ambitious projects such as the metaverse. 
However, XR is still limited by the current architecture of mobile systems. This paper makes the case for an XR-specific operating system (XROS). Such an XROS integrates hardware-support, computer vision algorithms, and XR-specific networking as the primitives supporting XR technology. These primitives represent the physical-digital world as a single shared resource among applications. Such an XROS allows for the development of coherent and system-wide interaction and display methods, systematic privacy preservation on sensor data, and performance improvement while simplifying application development. 


\end{abstract}

%
%



\maketitle

\section{Introduction}

Extended Reality (XR) and its applications are becoming increasingly pervasive. With the continually improving device hardware and the upgraded network capabilities brought by 5G, companies envision the development of large-scale XR environments. From the Niantic planet-scale AR alliance\footnote{\url{https://nianticlabs.com/blog/niantic-planet-scale-ar-alliance-5g}} to Facebook's metaverse, major industry actors envision the future of XR as pervasive, shared, and permanent environments where digital and physical are closely interleaved~\cite{lee2021needs}.


However, XR applications are currently far from such vision. XR presents tight performance constraints, with a minimum of 60\,FPS at high (4K) resolution and a motion-to-photon latency below 20\,ms. Current systems cannot achieve such performance, primarily owing to their architecture. Most platforms only allocate resources (e.g., camera) to a single application at a time, and handle XR processing in a compartmentalised fashion. As such, the digital reconstitution of the physical world, the content placement strategies, and the user interactions are unique to each application instance. Such strategy result in redundant development effort, limited performance, and prevents the integration of diverse applications within a single blended digital-physical world.

\begin{figure*}[t]
\centering
\begin{subfigure}[t]{0.6\textwidth}
\centering
\includegraphics[width=\textwidth]{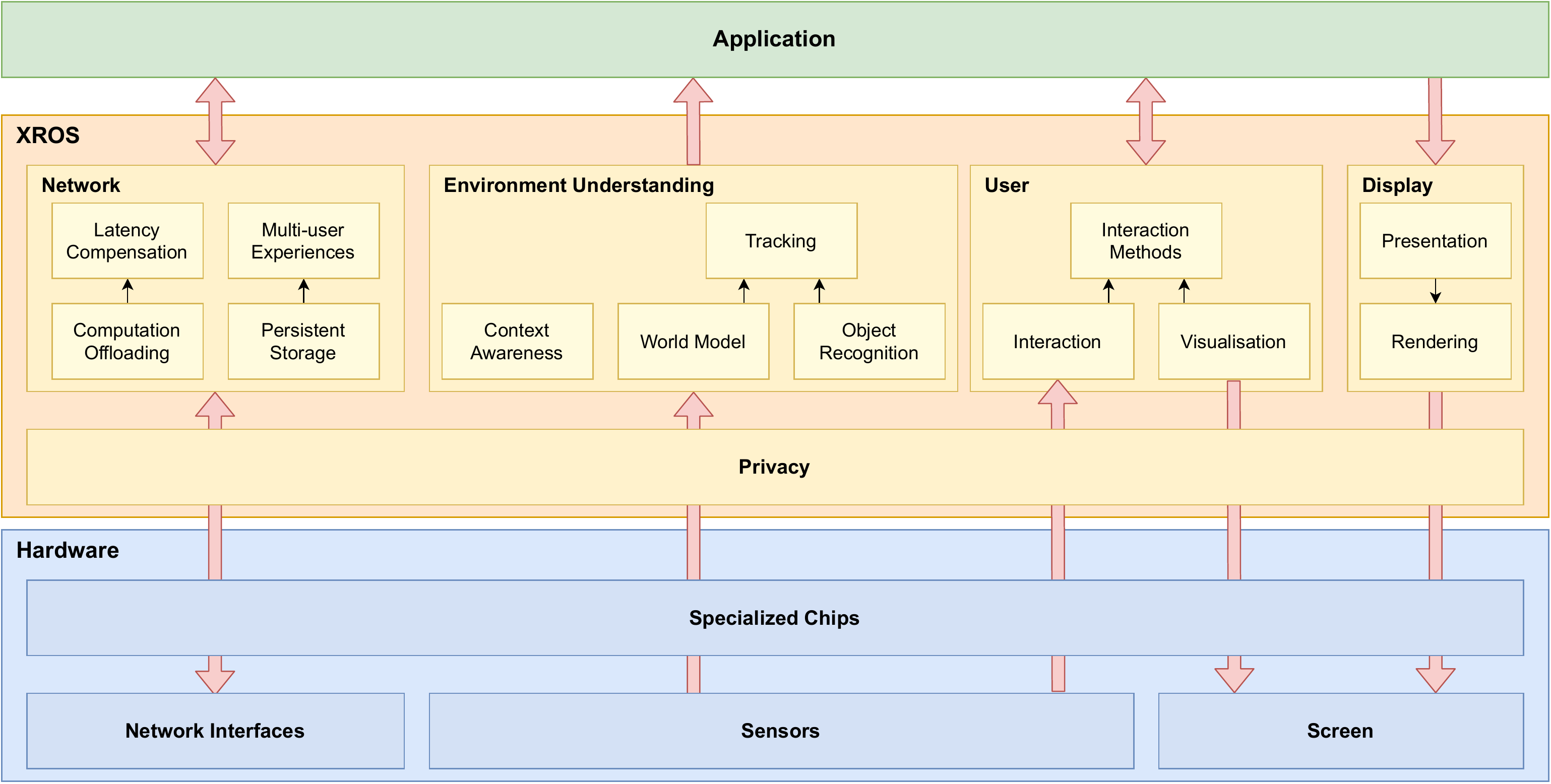}
\caption{Potential architecture of an XROS.}
\label{fig:archi}
\end{subfigure}\hfill
\begin{subfigure}[t]{0.39\textwidth}
\centering
\includegraphics[width=\textwidth]{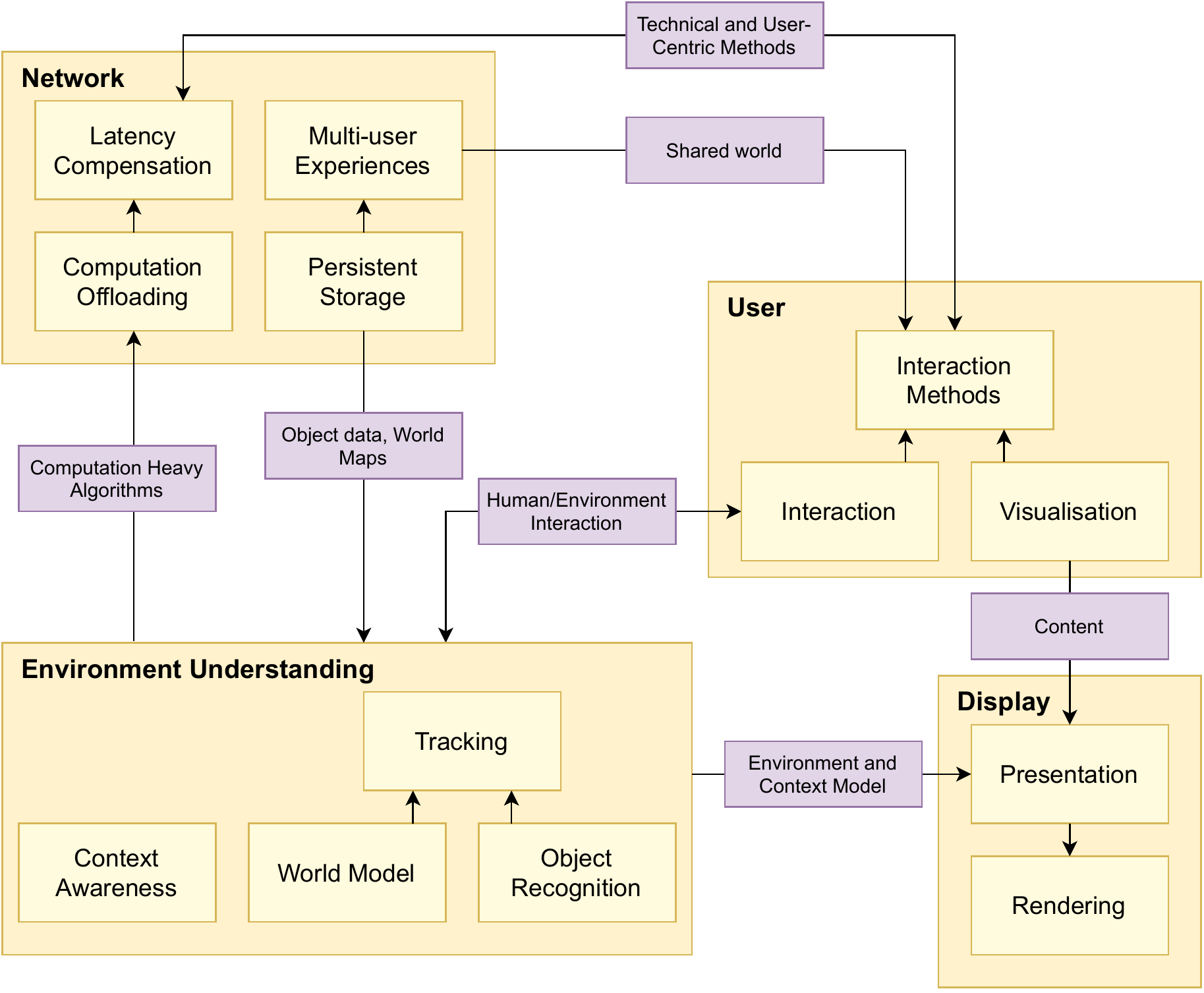}
\caption{Data flows between XROS's main modules. }
\label{fig:flows}
\end{subfigure}
\caption{XROS Architecture, major building blocks, and data flows.  Data coming from sensors is preprocessed by specialised chips and filtered by the privacy protection module before reaching the main building blocks of XR. These blocks provide the primitives for the XR application to execute in a standardised fashion.}
\end{figure*}

In this paper, we argue that the current model of application- or middleware- level XR is significantly impeding the development of XR applications, \edit{not to mention a metaverse. }
We propose for the enablers of XR to integrate the Operating System (OS) level. DiOS, our proposed architecture for an XROS, would integrate and expose XR fundamentals to provide (1) better performance by enabling fine-grained control of XR operation executions, (2) better integration between XR applications and the OS, and between XR applications, by considering the digital-physical XR world as a shared resource, (3) easier development of XR applications through standardised software primitives, (4) increased privacy and security by applying systematic privacy protection mechanisms on sensor data,  and (5) more intuitive and coherent user interaction through the usage of OS-level UIs. 




\section{Architecture of an XROS}

XR applications have stringent requirements that are difficult to achieve with current system architectures. 
By considering portions of the XR pipeline as resources to be shared between applications, the XROS enables such requirements. In this section, we first list XR application requirements, before detailing the architecture of the XROS.

\textbf{1. Requirements.}
XR applications superimpose digital content over the users' perception of the physical world. As such, their requirements are on par with the resolution of the users' perceptory system. Common guidelines cite a guaranteed minimum of \textit{60\,FPS, with 2k to 4K resolution, and a motion-to-photon latency below 20\,ms} for seamless XR experiences~\cite{8700215}. Such requirements are critical to prevent discrepancies (e.g., alignment) between the user's perception of the physical world and the digital content and preserve the immersion. 
Besides purely performance-related requirements, XR as an ubiquitous technology requires to \textit{share the physical-digital world model} between applications, and even users. As such, it makes sense for the model to be handled by the OS as a shared resource between applications, and devise user interaction methods coherent with such model.

\textbf{2. Architecture.}
An XROS enables the above requirements by (1) bringing computation closer to the hardware, (2) finely managing the execution timing of core XR operations, and (3) sharing the physical-digital world model among users and applications.  As such, the XROS moves from the current interruption-driven task execution model to a steady-state system where sensor data are processed under a periodic basis.
We propose an architecture revolving around six primary components as presented in Figure~\ref{fig:archi}: 

\textit{Environment Understanding:} XR applications build a representation of the users' surroundings through pervasive sensing.
It is necessary to integrate such techniques at OS-level to improve performance, enable XR in other OS modules, share the  physical world's model among services and applications, and provide standardised services to applications.

\textit{Specialised Chips Drivers:} Scene understanding relies on computation-heavy algorithms. A multitude of chips enable faster execution of such operations, from specialised chips to more general purpose elements such as tensor processing units. An XROS should leverage these chips to accelerate the execution of core elements of the pipeline.

\textit{Network:} Despite specialised hardware, some operations remain too intensive for mobile devices. It is thus necessary to offload these operations to more powerful remote servers. Shared and Persistent experiences also require the transmission and recombination of multiple individual perceptions of the environment towards building a global view of the perpetually evolving physical world. Ultra-low latency transmission will be a primary challenge of the XROS.


\textit{User Interaction:} A tight integration between Environment Understanding and User Interaction would enable novel interaction methods using the physical-digital duality of XR. Such synergy would allow users to leverage the physical world for interacting with digital content, whether embodied (e.g., gestures), or external (e.g., tangible interfaces).

\textit{Display:} Together with User Interaction, Display can benefit from the OS's Environment Understanding to blend digital content with the physical world while avoiding obstruction of critical physical objects (e.g., moving vehicles) or information overload in cluttered areas (e.g., busy streets).

\textit{Privacy:} To address the pervasive sensing of potentially private user and bystander information, a layer of privacy protection (whether software or hardware) should perform data minimisation between the sensors output and the other components' input. This layer would also further filter out data for network transmission and display.

Similar to mobile OSes based on UNIX, the remaining parts of the XROS (file system, process and memory management, networking) can be provided by open source components.

These components intercommunicate to reinforce integration of content with the physical-digital world, as shown in Figure~\ref{fig:flows}. The Environment Understanding module offloads computation-heavy operations through the Network module, which compensates latency together with the User Interaction module. The results from the Environment Understanding module are transmitted to the User Interaction  module to provide interaction methods that leverage the physical-digital duality of XR. Finally, the User Interaction module prepares the content for visualisation, which is combined with the world model from the Environment Understanding module to position content in the Display module. 

\section{Environment Understanding}

XR relies on pervasive environment sensing to 
 blend digital content with the physical world. In this section, we focus on the visual understanding of the physical world, reinforced by information from other sensors (depth cameras, LiDar)~\cite{campos2021orb}. To power XR applications and other OS modules, the environment understanding module should provide the base XR primitives and share the access to world model.

\textbf{1. XR Primitives.} An XROS should provide XR primitives to applications and services. Current OS-specific AR frameworks such as ARCore\footnote{\url{https://developers.google.com/ar}} or ARKit\footnote{\url{https://developer.apple.com/augmented-reality/}} 
combine sensor data to calculate and expose the 3D world model. They also provide functions such as plane identification, anchor recognition, and the possibility to save or share the experience. Although these frameworks provide a starting point for the XROS, none of them disclose how the feature points are identified, making the development of recognition and segmentation algorithms more difficult and redundant with the frameworks' algorithms. An XROS should explicitly expose the feature points extracted from the camera frames and how these points are refined through other sensors. 
The XROS would perform feature extraction and tracking and expose the results to every layer of the pipeline, while world building and object recognition could be implemented as standardised services to prevent redundancy and improve performance.

\textbf{2. Shared Access to the world's model.} One of the key functionalities of an operating system is to ensure the shared access to software and hardware resources for applications. The feature points, world model, semantic areas, and recognised objects are one of such resource. With a single physical world to share between applications, the OS needs to decide which application may display content over each area of the physical world~\cite{schmalstieg2002studierstube}. Such task does not only involve sharing the physical-digital world in-between applications, but also deciding which areas from the physical world may not be overlaid with content. The OS may define contextual rules  based on user safety, information density and visibility, or other cues driven by the user's habits.

\textbf{3. Specialised Hardware.} Integrating complex image and sensor processing operations in the OS may significantly impact the system's operation. Some operations are computation-heavy, while others rely on manipulating large data structures, and the pipeline should be processed under a steady-state basis. 
Specialised hardware may handle repetitive operations more efficiently. We consider three types of hardware: (1) highly specific single-operation chips for data preprocessing~\cite{nikolic2014synchronized}, (2) specialised hardware tapping into the memory of the processor to perform more complex operations~\cite{nguyen2019high}, and (3) less specialised hardware for executing general functions (rendering, machine learning, rendering). Nowadays, most mobile devices embed chips belonging to the third category. The XROS should leverage such chips to relieve the CPU load and minimise the impact of errors.
\section{Networking}

The previous section raised the need for on-device scene understanding to support XR applications. Remote scene understanding is also critical to execute XR operations that are either too computationally heavy, or to support persistent and shared experiences among users. The XROS would therefore be distributed in nature, supplementing standalone operation with data from nearby devices and the computation power of remote servers.

\begin{figure}[!t]
    \centering
    \includegraphics[width=\linewidth]{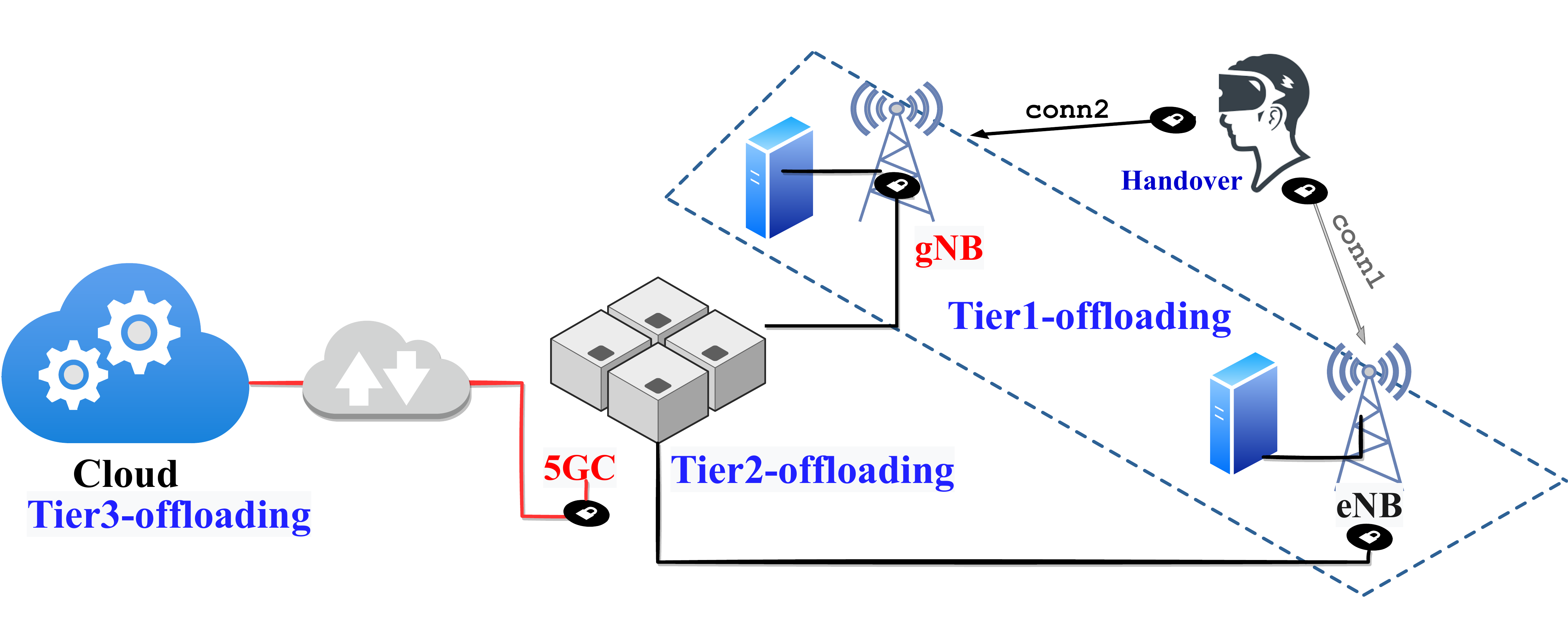}
    \caption{Handover and multi-tiers offloading in XROS.}
    \label{fig:networks}
\end{figure}

\textbf{1. Computation Offloading.}
The XROS may offload computation-heavy tasks to remote machines. Using background microservices, the XROS can estimate the computational cost of XR operations, maintain connections to remote servers,  provide handover on user's coverage change~\cite{tbraud2021talaria}, and decide when to offload tasks. It should also offer developers the possibility to specify whether a specific task should be executed on device or in the cloud~\cite{kosta2012thinkair}. 
Figure~\ref{fig:networks} illustrates how computation offloading and handover management could take place in 5G with a 2-tier server architecture. Background microservices detect when the user moves out of the eNB's range to the gNB, and  perform  handover and migration.

\textbf{2. Shared and Persistent Experiences.} One of the main projects for large-scale XR environments is the persistence of the XR world, allowing multiple users to share the same experience at the same location. 
Such applications aggregate the environment model of multiple devices to build a global model while enabling simultaneous interaction of multiple users with the digital content. The location-awareness of edge servers may be leveraged to perform this aggregation at a computationally acceptable scale~\cite{zhou2020edge}. Device-to-device communication may also be used for immediate communication between users~\cite{ansari20175g}. The XROS will thus face the challenge of relying on external computation providers for aggregating data and providing computation offloading.

\textbf{3. Content transmission.}
The XROS  will rely on distant servers, at the edge or in the cloud, to offload computations and provide shared experiences. XR presents significant bandwidth, latency and reliability constraints that need to be addressed.
Latency and bandwidth can be optimised through efficient usage of the available resources~\cite{braud2020multipath}. 
The transmission may experience packet loss owing to either congestion in the core network or channel error in the access network. An XROS should thus monitor the network conditions, and dynamically adapt the recovery techniques (retransmission or forward error correction). Similar to WebRTC, the XROS would obtain a better trade-off between temporal quality (smoothness of rendering), spatial video quality and end-to-end delay by utilising an adaptive hybrid NACK/FEC method~\cite{loss2013webrtc}. The XROS should also adapt the transmission rate according the available bandwidth. To this end, the XROS would either re-implement/redesign WebRTC (application protocol) to function as a transport protocol, or brings a novel protocols  that integrate adaptable video rate and efficient redundancy to mitigate packet losses.

\textbf{4. Latency Compensation.} Although computation offloading allows to execute sophisticated operations on constrained hardware, it comes at the cost of network latency. Such latency adds up to the rest of the pipeline, increasing the total motion-to-photon latency (the latency between user motion and its impact on the display), and thus reducing the immersion~\cite{braud2017future}. Similarly, shared experiences  Latency compensation techniques should be designed to hide such latency from the user. Services may  track the features of captured images until object recognition results are received to adjust to the device orientation~\cite{edgexar}. Geometric latency compensation may also be used to provide wider interaction targets in case of increased motion to photon.


\section{Novel Interactivity with XROS}\label{sec:user-inputs}

\begin{figure}[!t]
    \centering
    \includegraphics[width=\linewidth]{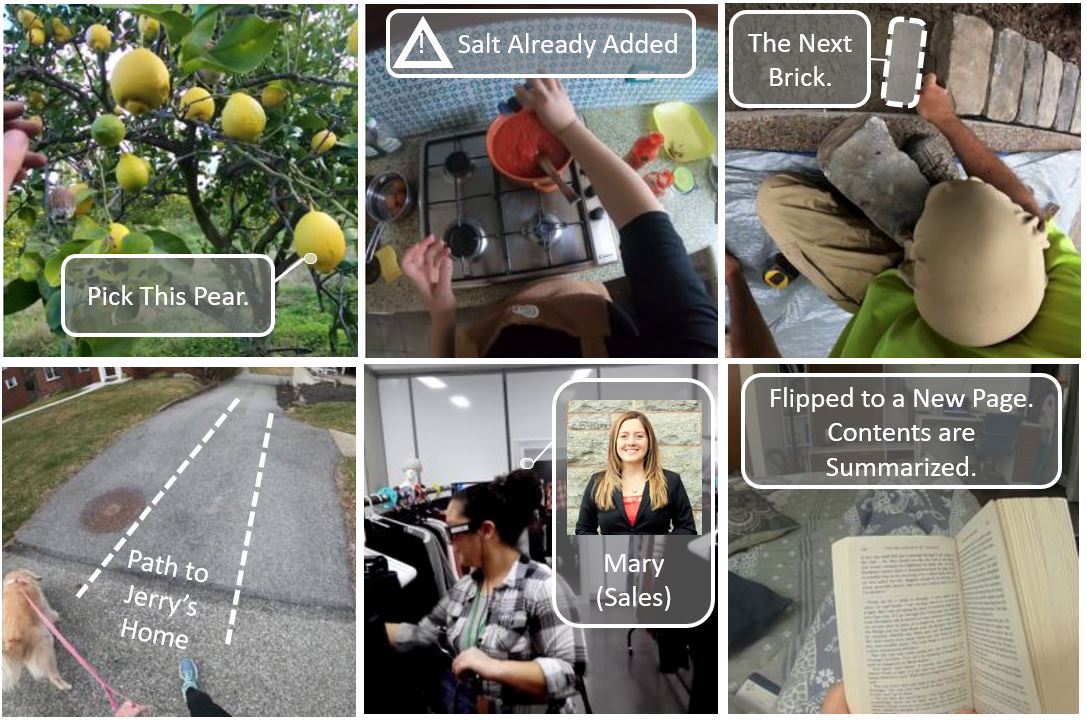}
    \caption{Virtual-Physical User Interaction in XROS\protect\footnotemark.}
    \label{fig:ego-act}
\end{figure}

The XROS goes beyond the 2D user interfaces dominated by desktops, windows, menus and icons, and emphasises the interactivity with users' physical bodies (e.g., gestures and bio-signal), surroundings (e.g., physical objects and bystanders), and other users~\cite{RBI-CHI08}.
Accordingly, we categorise the requirements of user interactivity as follows: 

\textbf{1. Blended UIs in the physical world.} 
The definition of augmented reality has been evolving from overlaying simple information to blended environments combining physical and digital seamlessly~\cite{MR-def-CHI19-Speicher}. Novel UIs across the digital and physical realities can serve numerous users to experience high levels of immersion and virtuality, in heterogeneous environments (Figure~\ref{fig:ego-act}). The XROS  should own the capacity of facilitating multiple users to collaborate with such virtual interfaces superimposed on the physical world. Such blended UIs should be ubiquitous, potentially up to city-scale~\cite{csur-lee-hcityI}. The challenges of achieving the city-wide requirements would need interdisciplinary efforts, primarily driven by a  \textit{distributed XROS} with \textit{advanced mobile networks} to synchronise multitudinous objects and their corresponding blended UIs through \textit{XR technology}. 




    \footnotetext{Images from the Ego4D dataset samples \url{https://ego4d-data.org/fig1.html}}

\textbf{2. Enhanced Awareness to Users and Their Environments.} 
To achieve blended UIs, the digital overlays should match with objects, locations and people in the physical world.
The XROS takes advantage of context awareness to understand the user's situated environment~\cite{cloud-or-not-cloud}. The XROS can achieve context awareness by recognising visual markers~\cite{edgexar} or 3D features on physical objects/users' body gestures~\cite{Hariharan2015HypercolumnsFO}. However, such recognition requires significant computational resources. Similar to the X windows system\footnote{https://x.org/wiki/}, the XROS may consider thin clients to display the overlay while  UI operations are performed on a separate machine. 

\textbf{3. Mobile User Interactivity.} 
The XROS  provides users with the fundamental input and output (I/O) functions. Users receive enriched AR information, whether visual, audio, or haptic~\cite{10.1145/3457141}). The key bottleneck of XR displays is the user's cognitive load. Displaying content without proper selection and management would cause information overflow and poor usability~\cite{Lindlbauer2019ContextAwareOA}.
Thus, the XROS has to maintain highly relevant content~\cite{A2W-kityung} inside the size-limited lens of XR~\cite{Lee2020FromST}.
The XROS should also adapt to the vastly diversified input interfaces, whether embodied or physical.


\textbf{4. Towards Tangible Interaction} As a final note, the blended UIs of the XROS do not limit to physical-digital mixed overlays. Tangible interaction will redefine our spatial environments. 
Remarkable examples of tangible interaction include  public displays, interactive tabletops, mechanised infrastructures, at-home IoT devices, drones, augmented surface of electronics~\cite{shaer2010tangible}. 
Thus, \textit{how to manage countless distinct tangible items} remains a critical challenge. 

Figure~\ref{fig:ego-act} presents several examples of user interaction  between digital and physical enabled by the XROS.

\section{Security and Privacy}

\begin{figure}
    \centering
    \includegraphics[width=\linewidth]{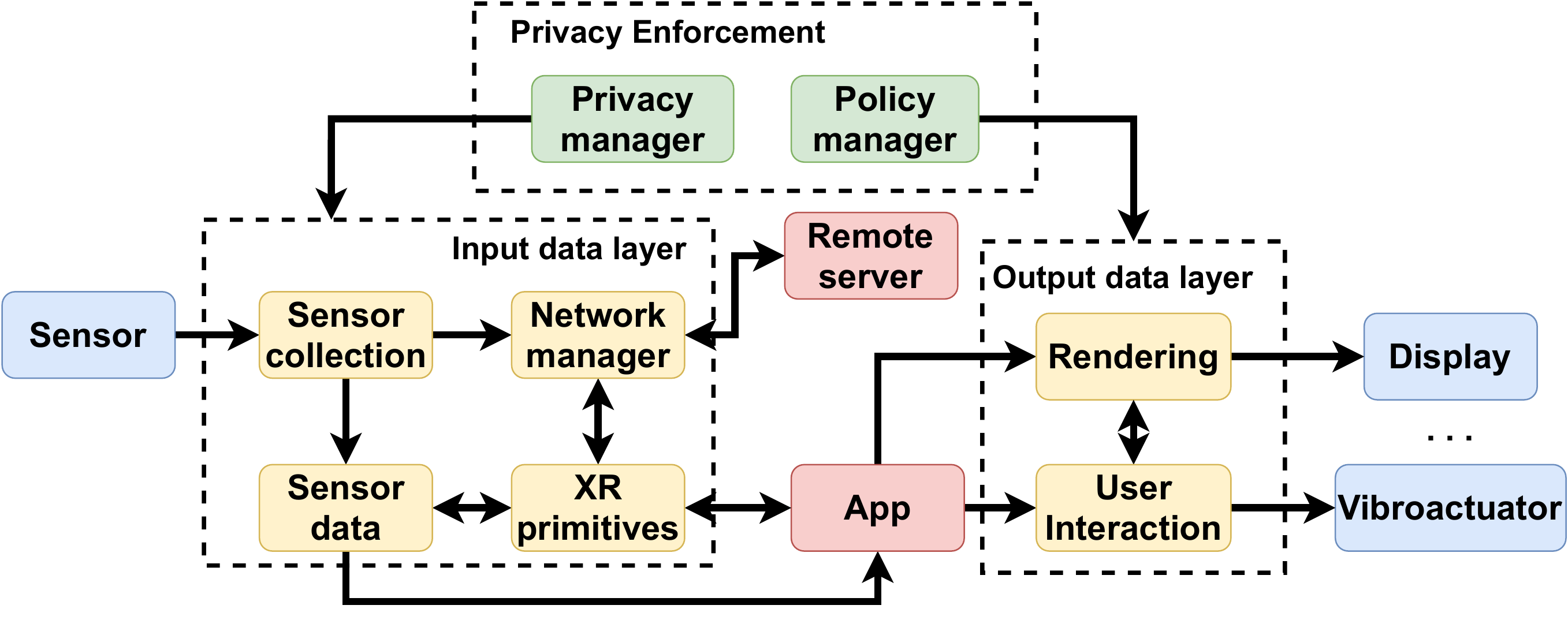}
    \caption{Privacy protection architecture filtering the application's input (sensor data) and output (display, haptic) flows with a privacy enforcement layer.}
    \label{fig:privact}
\end{figure}

XR operates through pervasively and continually sensing the users' physical surroundings. Therefore, it faces multiple security and privacy challenges depending on the technologies employed~\cite{roesner2014security}. In this section, we focus primarily on spatial data collection, whether on users or bystanders, and consider both the input and output flows as shown in Figure~\ref{fig:privact}.

\textbf{1. Spatial Data Collection}
Spatial data has become more ubiquitous with the raising adoption of XR technologies in mobile environments. This spatial information allow XR applications to acquire better and more precise understanding of the surroundings, but also opens new threats to users' and bystanders' privacy. Such data can be used to infer spaces from the captured 3D point clouds, and allow  an attacker to recognise space/objects belonging to a specific user despite the lack of visual images~\cite{guzman2021unravelling}. An XROS should take particular care to protect such data against potential threats.
\balance

\textbf{2. Bystander Privacy}
XR's pervasive sensing may affect the users' privacy. Many legal frameworks require data collection systems to give data subjects control over their personal data~\cite{EUdataregulations2018}. As non-users, bystanders often do not have such control, which complicates both the legal and ethical application of XR. An XROS should automatically detect and minimize data from bystanders~\cite{dimiccoli2018mitigating} to respect their privacy and adhere to legal frameworks.

\textbf{3. Input Data}
The XROS should include an input protection layer as an intermediary framework between raw sensors data and the XR applications. We build on top of~\cite{hu2021lenscap}, where the protection layer splits the access control and processing of the sensed data. This approach relies on three different modules: sensor collection, sensor data, and network manager. Each sensor includes a sensor collection module, where developers access the sensed data on-device. The sensed data is stored in the database module, where the applications can access and process the collected data. The network process allows the XR applications to share the sensed data with external parties (e.g., cloud computing, multiplayer game). Access to data by external parties (applications, remote server) is regulated by a privacy manager.

\textbf{4. Output Data and Safety}
Malicious XR output can raise new privacy threats~\cite{lebeck2017securing,roesner2014security}, for instance, displaying content over physical objects that present a risk for the user. Therefore,  the XROS includes a policy manager with a set of well-defined parameterized conditions and actions to perform in the XR output.
In~\cite{lebeck2017securing}, the authors propose an XR platform that controls the output that applications can display according to a set of context-aware policies. The output policies constrains the virtual content that users can see in the XR applications. The XROS policy manager should also handle  non-visual outputs such as audio and haptic, as well as the display priority when several applications are sharing the physical world's real estate. Besides safety issues, output data should undergo another layer of minimisation before being displayed or transmitted to remote machines.

\section{Conclusion}

\edit{This paper detailed the architecture of DiOS, an XROS architecture that addresses the stringent requirements of pervasive multi-users XR, enabling exciting developments such as the metaverse.}
The XROS integrates the main XR primitives at its core to provide other OS modules and applications with the base functions of XR. The XROS considers the physical world as a resource to be shared between applications, both in its operation and in terms of user interaction. Besides the vision algorithms driving XR functions, the XROS presents the particularity to be distributed in nature, to provide improved computation capabilities, and share the experience across multiple users. Finally, we detailed the privacy and security aspects of the XROS, to address the increasingly growing ethical and legal concerns with XR technology. 

\newpage

\bibliographystyle{ACM-Reference-Format}
\balance
\bibliography{references,privacy} 


\begin{thebibliography}{30}


\ifx \showCODEN    \undefined \def \showCODEN     #1{\unskip}     \fi
\ifx \showDOI      \undefined \def \showDOI       #1{#1}\fi
\ifx \showISBNx    \undefined \def \showISBNx     #1{\unskip}     \fi
\ifx \showISBNxiii \undefined \def \showISBNxiii  #1{\unskip}     \fi
\ifx \showISSN     \undefined \def \showISSN      #1{\unskip}     \fi
\ifx \showLCCN     \undefined \def \showLCCN      #1{\unskip}     \fi
\ifx \shownote     \undefined \def \shownote      #1{#1}          \fi
\ifx \showarticletitle \undefined \def \showarticletitle #1{#1}   \fi
\ifx \showURL      \undefined \def \showURL       {\relax}        \fi
\providecommand\bibfield[2]{#2}
\providecommand\bibinfo[2]{#2}
\providecommand\natexlab[1]{#1}
\providecommand\showeprint[2][]{arXiv:#2}

\bibitem[\protect\citeauthoryear{European Commission}{EUd}{[n.\,d.]}]%
        {EUdataregulations2018}
European Commission \bibinfo{year}{[n.\,d.]}\natexlab{}.
\newblock \bibinfo{booktitle}{\emph{2018 reform of EU data protection rules}}.
\newblock European Commission.
\newblock
\urldef\tempurl%
\url{https://ec.europa.eu/commission/sites/beta-political/files/data-protection-factsheet-changes_en.pdf}
\showURL{%
\tempurl}


\bibitem[\protect\citeauthoryear{Ansari, Chrysostomou, Hassan, Guizani, Mumtaz,
  Rodriguez, and Rodrigues}{Ansari et~al\mbox{.}}{2017}]%
        {ansari20175g}
\bibfield{author}{\bibinfo{person}{Rafay~Iqbal Ansari},
  \bibinfo{person}{Chrysostomos Chrysostomou}, \bibinfo{person}{Syed~Ali
  Hassan}, \bibinfo{person}{Mohsen Guizani}, \bibinfo{person}{Shahid Mumtaz},
  \bibinfo{person}{Jonathan Rodriguez}, {and} \bibinfo{person}{Joel~JPC
  Rodrigues}.} \bibinfo{year}{2017}\natexlab{}.
\newblock \showarticletitle{5G D2D networks: Techniques, challenges, and future
  prospects}.
\newblock \bibinfo{journal}{\emph{IEEE Systems Journal}} \bibinfo{volume}{12},
  \bibinfo{number}{4} (\bibinfo{year}{2017}), \bibinfo{pages}{3970--3984}.
\newblock


\bibitem[\protect\citeauthoryear{Bermejo, Lee, Chojecki, Przewozny, and
  Hui}{Bermejo et~al\mbox{.}}{2021}]%
        {10.1145/3457141}
\bibfield{author}{\bibinfo{person}{Carlos Bermejo}, \bibinfo{person}{Lik~Hang
  Lee}, \bibinfo{person}{Paul Chojecki}, \bibinfo{person}{David Przewozny},
  {and} \bibinfo{person}{Pan Hui}.} \bibinfo{year}{2021}\natexlab{}.
\newblock \showarticletitle{Exploring Button Designs for Mid-Air Interaction in
  Virtual Reality: A Hexa-Metric Evaluation of Key Representations and
  Multi-Modal Cues}.
\newblock \bibinfo{journal}{\emph{Proc. ACM Hum.-Comput. Interact.}}
  \bibinfo{volume}{5}, \bibinfo{number}{EICS}, Article \bibinfo{articleno}{194}
  (\bibinfo{date}{May} \bibinfo{year}{2021}), \bibinfo{numpages}{26}~pages.
\newblock


\bibitem[\protect\citeauthoryear{Braud, Alhilal, and Hui}{Braud
  et~al\mbox{.}}{2021}]%
        {tbraud2021talaria}
\bibfield{author}{\bibinfo{person}{Tristan Braud}, \bibinfo{person}{Ahmad
  Alhilal}, {and} \bibinfo{person}{Pan Hui}.} \bibinfo{year}{2021}\natexlab{}.
\newblock \showarticletitle{Talaria: In-engine Synchronisation for Seamless
  Migration of Mobile Edge Gaming Instances}. In
  \bibinfo{booktitle}{\emph{International Conference on emerging Networking
  EXperiments and Technologies (CoNEXT)}} (Munich, Germany).
\newblock


\bibitem[\protect\citeauthoryear{Braud, Bijarbooneh, Chatzopoulos, and
  Hui}{Braud et~al\mbox{.}}{2017}]%
        {braud2017future}
\bibfield{author}{\bibinfo{person}{Tristan Braud},
  \bibinfo{person}{Farshid~Hassani Bijarbooneh}, \bibinfo{person}{Dimitris
  Chatzopoulos}, {and} \bibinfo{person}{Pan Hui}.}
  \bibinfo{year}{2017}\natexlab{}.
\newblock \showarticletitle{Future networking challenges: The case of mobile
  augmented reality}. In \bibinfo{booktitle}{\emph{2017 IEEE 37th International
  Conference on Distributed Computing Systems (ICDCS)}}. IEEE,
  \bibinfo{pages}{1796--1807}.
\newblock


\bibitem[\protect\citeauthoryear{Braud, Zhou, Kangasharju, and Hui}{Braud
  et~al\mbox{.}}{2020}]%
        {braud2020multipath}
\bibfield{author}{\bibinfo{person}{Tristan Braud}, \bibinfo{person}{Pengyuan
  Zhou}, \bibinfo{person}{Jussi Kangasharju}, {and} \bibinfo{person}{Pan Hui}.}
  \bibinfo{year}{2020}\natexlab{}.
\newblock \showarticletitle{Multipath computation offloading for mobile
  augmented reality}. In \bibinfo{booktitle}{\emph{2020 IEEE International
  Conference on Pervasive Computing and Communications (PerCom)}}. IEEE,
  \bibinfo{pages}{1--10}.
\newblock


\bibitem[\protect\citeauthoryear{Campos, Elvira, Rodr{\'\i}guez, Montiel, and
  Tard{\'o}s}{Campos et~al\mbox{.}}{2021}]%
        {campos2021orb}
\bibfield{author}{\bibinfo{person}{Carlos Campos}, \bibinfo{person}{Richard
  Elvira}, \bibinfo{person}{Juan J~G{\'o}mez Rodr{\'\i}guez},
  \bibinfo{person}{Jos{\'e}~MM Montiel}, {and} \bibinfo{person}{Juan~D
  Tard{\'o}s}.} \bibinfo{year}{2021}\natexlab{}.
\newblock \showarticletitle{ORB-SLAM3: An Accurate Open-Source Library for
  Visual, Visual--Inertial, and Multimap SLAM}.
\newblock \bibinfo{journal}{\emph{IEEE Transactions on Robotics}}
  (\bibinfo{year}{2021}).
\newblock


\bibitem[\protect\citeauthoryear{Dimiccoli, Mar{\'\i}n, and Thomaz}{Dimiccoli
  et~al\mbox{.}}{2018}]%
        {dimiccoli2018mitigating}
\bibfield{author}{\bibinfo{person}{Mariella Dimiccoli}, \bibinfo{person}{Juan
  Mar{\'\i}n}, {and} \bibinfo{person}{Edison Thomaz}.}
  \bibinfo{year}{2018}\natexlab{}.
\newblock \showarticletitle{Mitigating bystander privacy concerns in egocentric
  activity recognition with deep learning and intentional image degradation}.
\newblock \bibinfo{journal}{\emph{Proc. of the ACM on IMWUT}}
  \bibinfo{volume}{1}, \bibinfo{number}{4} (\bibinfo{year}{2018}),
  \bibinfo{pages}{1--18}.
\newblock


\bibitem[\protect\citeauthoryear{Guzman, Seneviratne, and Thilakarathna}{Guzman
  et~al\mbox{.}}{2021}]%
        {guzman2021unravelling}
\bibfield{author}{\bibinfo{person}{Jaybie Agullo~de Guzman},
  \bibinfo{person}{Aruna Seneviratne}, {and} \bibinfo{person}{Kanchana
  Thilakarathna}.} \bibinfo{year}{2021}\natexlab{}.
\newblock \showarticletitle{Unravelling Spatial Privacy Risks of Mobile Mixed
  Reality Data}.
\newblock \bibinfo{journal}{\emph{Proceedings of the ACM on Interactive,
  Mobile, Wearable and Ubiquitous Technologies (IMWUT)}} \bibinfo{volume}{5},
  \bibinfo{number}{1} (\bibinfo{year}{2021}), \bibinfo{pages}{1--26}.
\newblock


\bibitem[\protect\citeauthoryear{Hariharan, Arbel{\'a}ez, Girshick, and
  Malik}{Hariharan et~al\mbox{.}}{2015}]%
        {Hariharan2015HypercolumnsFO}
\bibfield{author}{\bibinfo{person}{Bharath Hariharan}, \bibinfo{person}{Pablo
  Arbel{\'a}ez}, \bibinfo{person}{Ross~B. Girshick}, {and}
  \bibinfo{person}{Jitendra Malik}.} \bibinfo{year}{2015}\natexlab{}.
\newblock \showarticletitle{Hypercolumns for object segmentation and
  fine-grained localization}.
\newblock \bibinfo{journal}{\emph{2015 IEEE Conference on Computer Vision and
  Pattern Recognition (CVPR)}} (\bibinfo{year}{2015}),
  \bibinfo{pages}{447--456}.
\newblock


\bibitem[\protect\citeauthoryear{Holmer, Shemer, and Paniconi}{Holmer
  et~al\mbox{.}}{2013}]%
        {loss2013webrtc}
\bibfield{author}{\bibinfo{person}{Stefan Holmer}, \bibinfo{person}{Mikhal
  Shemer}, {and} \bibinfo{person}{Marco Paniconi}.}
  \bibinfo{year}{2013}\natexlab{}.
\newblock \showarticletitle{Handling packet loss in WebRTC}. In
  \bibinfo{booktitle}{\emph{2013 IEEE International Conference on Image
  Processing}}. \bibinfo{pages}{1860--1864}.
\newblock


\bibitem[\protect\citeauthoryear{Hu, Iosifescu, and LiKamWa}{Hu
  et~al\mbox{.}}{2021}]%
        {hu2021lenscap}
\bibfield{author}{\bibinfo{person}{Jinhan Hu}, \bibinfo{person}{Andrei
  Iosifescu}, {and} \bibinfo{person}{Robert LiKamWa}.}
  \bibinfo{year}{2021}\natexlab{}.
\newblock \showarticletitle{LensCap: split-process framework for fine-grained
  visual privacy control for augmented reality apps}. In
  \bibinfo{booktitle}{\emph{Proceedings of the 19th Annual International
  Conference on Mobile Systems, Applications, and Services}}.
  \bibinfo{pages}{14--27}.
\newblock


\bibitem[\protect\citeauthoryear{Jacob, Girouard, Hirshfield, Horn, Shaer,
  Solovey, and Zigelbaum}{Jacob et~al\mbox{.}}{2008}]%
        {RBI-CHI08}
\bibfield{author}{\bibinfo{person}{Robert~J.K. Jacob}, \bibinfo{person}{Audrey
  Girouard}, \bibinfo{person}{Leanne~M. Hirshfield},
  \bibinfo{person}{Michael~S. Horn}, \bibinfo{person}{Orit Shaer},
  \bibinfo{person}{Erin~Treacy Solovey}, {and} \bibinfo{person}{Jamie
  Zigelbaum}.} \bibinfo{year}{2008}\natexlab{}.
\newblock \showarticletitle{Reality-Based Interaction: A Framework for
  Post-WIMP Interfaces}. In \bibinfo{booktitle}{\emph{Proc. of the SIGCHI
  Conference on Human Factors in Computing Systems}} (Florence, Italy).
  \bibinfo{pages}{201–210}.
\newblock
\showISBNx{9781605580111}


\bibitem[\protect\citeauthoryear{Kosta, Aucinas, Hui, Mortier, and Zhang}{Kosta
  et~al\mbox{.}}{2012}]%
        {kosta2012thinkair}
\bibfield{author}{\bibinfo{person}{Sokol Kosta}, \bibinfo{person}{Andrius
  Aucinas}, \bibinfo{person}{Pan Hui}, \bibinfo{person}{Richard Mortier}, {and}
  \bibinfo{person}{Xinwen Zhang}.} \bibinfo{year}{2012}\natexlab{}.
\newblock \showarticletitle{Thinkair: Dynamic resource allocation and parallel
  execution in the cloud for mobile code offloading}. In
  \bibinfo{booktitle}{\emph{2012 Proc. IEEE Infocom}}. IEEE,
  \bibinfo{pages}{945--953}.
\newblock


\bibitem[\protect\citeauthoryear{Lai, Hu, Cui, Sun, Dai, and Lee}{Lai
  et~al\mbox{.}}{2020}]%
        {8700215}
\bibfield{author}{\bibinfo{person}{Zeqi Lai}, \bibinfo{person}{Y.~Charlie Hu},
  \bibinfo{person}{Yong Cui}, \bibinfo{person}{Linhui Sun},
  \bibinfo{person}{Ningwei Dai}, {and} \bibinfo{person}{Hung-Sheng Lee}.}
  \bibinfo{year}{2020}\natexlab{}.
\newblock \showarticletitle{Furion: Engineering High-Quality Immersive Virtual
  Reality on Today's Mobile Devices}.
\newblock \bibinfo{journal}{\emph{IEEE Transactions on Mobile Computing}}
  \bibinfo{volume}{19}, \bibinfo{number}{7} (\bibinfo{year}{2020}),
  \bibinfo{pages}{1586--1602}.
\newblock


\bibitem[\protect\citeauthoryear{Lam, Lee, and Hui}{Lam et~al\mbox{.}}{2021}]%
        {A2W-kityung}
\bibfield{author}{\bibinfo{person}{Kit~Yung Lam}, \bibinfo{person}{Lik~Hang
  Lee}, {and} \bibinfo{person}{Pan Hui}.} \bibinfo{year}{2021}\natexlab{}.
\newblock \showarticletitle{A2W: Context-Aware Recommendation System for Mobile
  Augmented Reality Web Browser}. In \bibinfo{booktitle}{\emph{Proc. of the
  29th ACM International Conference on Multimedia}} (Virtual Event, China)
  \emph{(\bibinfo{series}{MM '21})}. \bibinfo{pages}{2447–2455}.
\newblock
\showISBNx{9781450386517}


\bibitem[\protect\citeauthoryear{Lebeck, Ruth, Kohno, and Roesner}{Lebeck
  et~al\mbox{.}}{2017}]%
        {lebeck2017securing}
\bibfield{author}{\bibinfo{person}{Kiron Lebeck}, \bibinfo{person}{Kimberly
  Ruth}, \bibinfo{person}{Tadayoshi Kohno}, {and} \bibinfo{person}{Franziska
  Roesner}.} \bibinfo{year}{2017}\natexlab{}.
\newblock \showarticletitle{Securing augmented reality output}. In
  \bibinfo{booktitle}{\emph{2017 IEEE symposium on security and privacy (SP)}}.
  IEEE, \bibinfo{pages}{320--337}.
\newblock


\bibitem[\protect\citeauthoryear{Lee, Braud, Hosio, and Hui}{Lee
  et~al\mbox{.}}{2021a}]%
        {csur-lee-hcityI}
\bibfield{author}{\bibinfo{person}{Lik-Hang Lee}, \bibinfo{person}{Tristan
  Braud}, \bibinfo{person}{Simo Hosio}, {and} \bibinfo{person}{Pan Hui}.}
  \bibinfo{year}{2021}\natexlab{a}.
\newblock \showarticletitle{Towards Augmented Reality Driven Human-City
  Interaction: Current Research on Mobile Headsets and Future Challenges}.
\newblock \bibinfo{journal}{\emph{ACM Comput. Surv.}} \bibinfo{volume}{54},
  \bibinfo{number}{8}, Article \bibinfo{articleno}{165} (\bibinfo{date}{Oct.}
  \bibinfo{year}{2021}), \bibinfo{numpages}{38}~pages.
\newblock
\showISSN{0360-0300}
\urldef\tempurl%
\url{https://doi.org/10.1145/3467963}
\showDOI{\tempurl}


\bibitem[\protect\citeauthoryear{Lee, Braud, Lam, Yau, and Hui}{Lee
  et~al\mbox{.}}{2020}]%
        {Lee2020FromST}
\bibfield{author}{\bibinfo{person}{Lik-Hang Lee}, \bibinfo{person}{Tristan
  Braud}, \bibinfo{person}{Kit-Yung Lam}, \bibinfo{person}{Yui-Pan Yau}, {and}
  \bibinfo{person}{Pan Hui}.} \bibinfo{year}{2020}\natexlab{}.
\newblock \showarticletitle{From seen to unseen: Designing keyboard-less
  interfaces for text entry on the constrained screen real estate of Augmented
  Reality headsets}.
\newblock \bibinfo{journal}{\emph{Pervasive Mob. Comput.}}
  \bibinfo{volume}{64} (\bibinfo{year}{2020}), \bibinfo{pages}{101148}.
\newblock


\bibitem[\protect\citeauthoryear{Lee, Braud, Zhou, Wang, Xu, Lin, Kumar,
  Bermejo, and Hui}{Lee et~al\mbox{.}}{2021b}]%
        {lee2021needs}
\bibfield{author}{\bibinfo{person}{Lik-Hang Lee}, \bibinfo{person}{Tristan
  Braud}, \bibinfo{person}{Pengyuan Zhou}, \bibinfo{person}{Lin Wang},
  \bibinfo{person}{Dianlei Xu}, \bibinfo{person}{Zijun Lin},
  \bibinfo{person}{Abhishek Kumar}, \bibinfo{person}{Carlos Bermejo}, {and}
  \bibinfo{person}{Pan Hui}.} \bibinfo{year}{2021}\natexlab{b}.
\newblock \bibinfo{title}{All One Needs to Know about Metaverse: A Complete
  Survey on Technological Singularity, Virtual Ecosystem, and Research Agenda}.
\newblock
\newblock
\showeprint[arxiv]{2110.05352}~[cs.CY]


\bibitem[\protect\citeauthoryear{Lindlbauer, Feit, and Hilliges}{Lindlbauer
  et~al\mbox{.}}{2019}]%
        {Lindlbauer2019ContextAwareOA}
\bibfield{author}{\bibinfo{person}{David Lindlbauer},
  \bibinfo{person}{Anna~Maria Feit}, {and} \bibinfo{person}{Otmar Hilliges}.}
  \bibinfo{year}{2019}\natexlab{}.
\newblock \showarticletitle{Context-Aware Online Adaptation of Mixed Reality
  Interfaces}.
\newblock \bibinfo{journal}{\emph{Proc. of the 32nd Annual ACM Symp. on UIST}}
  (\bibinfo{year}{2019}).
\newblock


\bibitem[\protect\citeauthoryear{Naqvi, Moens, Ramakrishnan, Preuveneers,
  Hughes, and Berbers}{Naqvi et~al\mbox{.}}{2015}]%
        {cloud-or-not-cloud}
\bibfield{author}{\bibinfo{person}{Nayyab~Zia Naqvi}, \bibinfo{person}{Karel
  Moens}, \bibinfo{person}{Arun Ramakrishnan}, \bibinfo{person}{Davy
  Preuveneers}, \bibinfo{person}{Danny Hughes}, {and} \bibinfo{person}{Yolande
  Berbers}.} \bibinfo{year}{2015}\natexlab{}.
\newblock \showarticletitle{To Cloud or Not to Cloud: A Context-Aware
  Deployment Perspective of Augmented Reality Mobile Applications}. In
  \bibinfo{booktitle}{\emph{Proc. of the 30th Annual ACM Symp. on Applied
  Computing}} (Salamanca, Spain) \emph{(\bibinfo{series}{SAC '15})}.
  \bibinfo{pages}{555–562}.
\newblock
\showISBNx{9781450331968}


\bibitem[\protect\citeauthoryear{Nguyen, Nguyen, Kim, and Lee}{Nguyen
  et~al\mbox{.}}{2019}]%
        {nguyen2019high}
\bibfield{author}{\bibinfo{person}{Duy~Thanh Nguyen},
  \bibinfo{person}{Tuan~Nghia Nguyen}, \bibinfo{person}{Hyun Kim}, {and}
  \bibinfo{person}{Hyuk-Jae Lee}.} \bibinfo{year}{2019}\natexlab{}.
\newblock \showarticletitle{A high-throughput and power-efficient FPGA
  implementation of YOLO CNN for object detection}.
\newblock \bibinfo{journal}{\emph{IEEE Transactions on Very Large Scale
  Integration (VLSI) Systems}} \bibinfo{volume}{27}, \bibinfo{number}{8}
  (\bibinfo{year}{2019}), \bibinfo{pages}{1861--1873}.
\newblock


\bibitem[\protect\citeauthoryear{Nikolic, Rehder, Burri, Gohl, Leutenegger,
  Furgale, and Siegwart}{Nikolic et~al\mbox{.}}{2014}]%
        {nikolic2014synchronized}
\bibfield{author}{\bibinfo{person}{Janosch Nikolic}, \bibinfo{person}{Joern
  Rehder}, \bibinfo{person}{Michael Burri}, \bibinfo{person}{Pascal Gohl},
  \bibinfo{person}{Stefan Leutenegger}, \bibinfo{person}{Paul~T Furgale}, {and}
  \bibinfo{person}{Roland Siegwart}.} \bibinfo{year}{2014}\natexlab{}.
\newblock \showarticletitle{A synchronized visual-inertial sensor system with
  FPGA pre-processing for accurate real-time SLAM}. In
  \bibinfo{booktitle}{\emph{2014 IEEE international conference on robotics and
  automation (ICRA)}}. IEEE, \bibinfo{pages}{431--437}.
\newblock


\bibitem[\protect\citeauthoryear{Roesner, Kohno, and Molnar}{Roesner
  et~al\mbox{.}}{2014}]%
        {roesner2014security}
\bibfield{author}{\bibinfo{person}{Franziska Roesner},
  \bibinfo{person}{Tadayoshi Kohno}, {and} \bibinfo{person}{David Molnar}.}
  \bibinfo{year}{2014}\natexlab{}.
\newblock \showarticletitle{Security and privacy for augmented reality
  systems}.
\newblock \bibinfo{journal}{\emph{Comm. of the ACM}} \bibinfo{volume}{57},
  \bibinfo{number}{4} (\bibinfo{year}{2014}), \bibinfo{pages}{88--96}.
\newblock


\bibitem[\protect\citeauthoryear{Schmalstieg, Fuhrmann, Hesina, Szalav{\'a}ri,
  Encarna{\c{c}}ao, Gervautz, and Purgathofer}{Schmalstieg
  et~al\mbox{.}}{2002}]%
        {schmalstieg2002studierstube}
\bibfield{author}{\bibinfo{person}{Dieter Schmalstieg}, \bibinfo{person}{Anton
  Fuhrmann}, \bibinfo{person}{Gerd Hesina}, \bibinfo{person}{Zsolt
  Szalav{\'a}ri}, \bibinfo{person}{L~Miguel Encarna{\c{c}}ao},
  \bibinfo{person}{Michael Gervautz}, {and} \bibinfo{person}{Werner
  Purgathofer}.} \bibinfo{year}{2002}\natexlab{}.
\newblock \showarticletitle{The studierstube augmented reality project}.
\newblock \bibinfo{journal}{\emph{Presence: Teleoperators \& Virtual
  Environments}} \bibinfo{volume}{11}, \bibinfo{number}{1}
  (\bibinfo{year}{2002}), \bibinfo{pages}{33--54}.
\newblock


\bibitem[\protect\citeauthoryear{Shaer and Hornecker}{Shaer and
  Hornecker}{2010}]%
        {shaer2010tangible}
\bibfield{author}{\bibinfo{person}{Orit Shaer} {and} \bibinfo{person}{Eva
  Hornecker}.} \bibinfo{year}{2010}\natexlab{}.
\newblock \bibinfo{booktitle}{\emph{Tangible user interfaces: past, present,
  and future directions}}.
\newblock \bibinfo{publisher}{Now Publishers Inc}.
\newblock


\bibitem[\protect\citeauthoryear{Speicher, Hall, and Nebeling}{Speicher
  et~al\mbox{.}}{2019}]%
        {MR-def-CHI19-Speicher}
\bibfield{author}{\bibinfo{person}{Maximilian Speicher},
  \bibinfo{person}{Brian~D. Hall}, {and} \bibinfo{person}{Michael Nebeling}.}
  \bibinfo{year}{2019}\natexlab{}.
\newblock \showarticletitle{What is Mixed Reality?}. In
  \bibinfo{booktitle}{\emph{Proc. of the 2019 CHI Conference on Human Factors
  in Computing Systems}}. \bibinfo{pages}{1–15}.
\newblock
\showISBNx{9781450359702}


\bibitem[\protect\citeauthoryear{Zhang, Lin, Bijarbooneh, Cheng, Braud, Zhou,
  Lee, and Hui}{Zhang et~al\mbox{.}}{2022}]%
        {edgexar}
\bibfield{author}{\bibinfo{person}{Wenxiao Zhang}, \bibinfo{person}{Sikun Lin},
  \bibinfo{person}{Farshid Bijarbooneh}, \bibinfo{person}{Hao-Fei Cheng},
  \bibinfo{person}{Tristan Braud}, \bibinfo{person}{Pengyuan Zhou},
  \bibinfo{person}{Lik-Hang Lee}, {and} \bibinfo{person}{Pan Hui}.}
  \bibinfo{year}{2022}\natexlab{}.
\newblock \showarticletitle{EdgeXAR: A 6-DoF Camera Multi-target Interaction
  Framework for MAR with User-friendly Latency Compensation using Edge
  Computing}. In \bibinfo{booktitle}{\emph{Proc. of the ACM on HCI (EICS)}}.
\newblock


\bibitem[\protect\citeauthoryear{Zhou, Braud, Zavodovski, Liu, Chen, Hui, and
  Kangasharju}{Zhou et~al\mbox{.}}{2020}]%
        {zhou2020edge}
\bibfield{author}{\bibinfo{person}{Pengyuan Zhou}, \bibinfo{person}{Tristan
  Braud}, \bibinfo{person}{Aleksandr Zavodovski}, \bibinfo{person}{Zhi Liu},
  \bibinfo{person}{Xianfu Chen}, \bibinfo{person}{Pan Hui}, {and}
  \bibinfo{person}{Jussi Kangasharju}.} \bibinfo{year}{2020}\natexlab{}.
\newblock \showarticletitle{Edge-facilitated augmented vision in
  vehicle-to-everything networks}.
\newblock \bibinfo{journal}{\emph{IEEE Transactions on Vehicular Technology}}
  \bibinfo{volume}{69}, \bibinfo{number}{10} (\bibinfo{year}{2020}),
  \bibinfo{pages}{12187--12201}.
\newblock


\end{thebibliography}

\end{document}